\documentclass[twocolumn]{aa}
\usepackage{graphicx}
\usepackage{xspace}
\usepackage{subcaption}
\usepackage{lmodern}
\usepackage{anyfontsize}
\RequirePackage{fix-cm}
\usepackage[OML,OMS,OT2,OT1]{fontenc}
\usepackage[]{fontenc}
\usepackage{natbib}

\usepackage{txfonts}
\usepackage{amsmath}
\usepackage{xcolor}
\usepackage[title]{appendix}

\graphicspath{{./}{figures/}}
\begin{document}

\title{Hydrodynamical  simulations of protoplanetary disks including irradiation of stellar photons}

\subtitle{I. Resolution study for Vertical Shear Instability (VSI)}

\author{L. Flores-Rivera
          \inst{1}
          M. Flock\inst{1}
          \and
          R. Nakatani\inst{2}\fnmsep
          }

\institute{Max-Planck Institute for Astronomy, K\"onigstuhl 17, 69117 Heidelberg, Germany\\
              \email{flores@mpia.de} \\
              \email{flock@mpia.de}
         \and
             2-1, Hirosawa, Wako City, Saitama Pref. No. S407, 4F, Materials Science Research Building, RIKEN \\
             \email{ryohei.nakatani@riken.jp}
             }


 
  \abstract
   {In recent years hydrodynamical (HD) models have become important to describe the gas kinematics in protoplanetary disks,  especially in combination with models of photoevaporation and/or magnetic-driven winds. Our aim is to investigate how the Vertical Shear Instability (VSI) could influence the thermally-driven winds on the surface of protoplanetary disks.}
   {In this first part of the project, we focus on diagnosing the conditions of the VSI at the highest numerical resolution ever recorded and allude at what resolution per scale height we obtain convergence. At the same time, we want to investigate the vertical extent of VSI activity. Finally, we determine the regions where EUV, FUV and X-Rays are dominant in the disk.}
   {We perform global HD simulations using the \textsc{PLUTO} code. We adopt a global isothermal accretion disk setup, 2.5D (2 dimensions, 3 components) which covers a radial domain from 0.5 to 5.0 and an approximately full meridional extension. Our simulation runs cover a resolution from 12 to 203 cells per scale height.} 
   {We determine the 50 cells per scale height to be the lower limit to resolve the VSI. For higher resolutions, $\geq$50 cells per scale height, we observe the convergence for the saturation level of the kinetic energy.  We are also able to identify the growth of the `body' modes, with higher growth rate for higher resolution. Full energy saturation and a turbulent steady state is reached after 70 local orbits. We determine the location of the EUV-heated region defined by $\Sigma_{r}$=10$^{19}$ cm$^{-2}$ to be at $H_\mathrm{R}\sim9.7$ and the FUV/X-Rays-heated boundary layer defined by $\Sigma_{r}$=10$^{22}$ cm$^{-2}$ to be at $H_\mathrm{R}\sim6.2$, making it necessary to introduce the need of a hot atmosphere. For the first time, we report the presence of small scale vortices in the $r-Z$ plane, between the characteristic layers of large scale vertical velocity motions. Such vortices could lead to dust concentration, promoting grain growth. Our results highlight the importance to combine photoevaporation processes in the future high-resolution studies of the turbulence and accretion processes in disks. }
   {}
  %


\titlerunning{Vertical Shear Instability and Photoevaporative Winds}
\authorrunning{L. Flores-Rivea, M. Flock, \& R. Nakatani}
    
   \maketitle

\section{Introduction}

Hydrodynamical (HD) simulations in protoplanetary disks have gained substantial attention during the last decades. Due to the low coupling between the magnetic field and the gas in a large fraction of the disk \citep{Turner_2014, Dzyurkevich_2013} we expect that HD instabilities play a more important role for the gas kinematics and the overall gas evolution in protoplanetary disks. In theory, the presence of the VSI \citep{Goldreich_1967, Fricke_1968} in purely HD accretion disks has been well studied by careful perturbational analysis and numerical simulations in several works (e.g \citet{Nelson_2013, 2014A&A...572A..77S, Barker_2015, Lin_2015}). Several works have shown the importance of this instability to explain the turbulence and angular momentum transport in the disk \citep{arlt2004simulations, Nelson_2013, 2014A&A...572A..77S}. The main ingredients of the VSI are a short thermal relaxation timescale and a vertical gradient of the rotational velocity in the disk. However, understanding the effect of the VSI together with realistic thermal profiles in the disk are not well known. The evolution of the VSI has being tested in the context of planet formation and disk evolution. Recent simulations conducted by \citet{Lin2019} have found the VSI can be suppressed by dust settling and grain growth, however, the Streaming Instability (SI) may be present when VSI is active in the turbulent disk \citep{2020A&A...635A.190S}. In the context of disk evolution, its been proposed (e.g., \citet{Shu_1994, Hollenbach_1994, Clarke_2001, Owen_2012, Alexander_2006I, Alexander_2006, Alexander_2014, Ercolano_2017}) that photoevaporation by energetic photons is a potential agent of disk dispersal. Investigating the effect of high energy radiation onto protoplanetary disks can be quite complex, but recent efforts (e.g., \citet{Ercolano_2008, Gorti_2009, Owen_2010, Owen_2011}) have attempted to explain the effect of high energy radiation onto the disk through observations of winds. A prominent signature of photoevaporation processes is the detection of forbidden lines in the protoplanetary disk spectrum. The detection of NeII line from the surface of TW Hya at around 10 AU \citep{Pascucci_2011}, shows the important influence of high energetic photons onto the surface layers of the disk, however, the vertical extension of the wind is quite inconclusive yet. Moreover, \citet{Ballabio_2020} provided synthetic line profile of NeII to be consistent with thermally driven winds while other forbidden molecules such as OI, OII and SII require a different scenario, possible magneto driven winds \citep{Fang2018,Banzatti2019}. Photoevaporative process in the inner disk (< 5 AU) has been tested by \citet{Wang_2019} and \citet{Gressel_2020} which considered hydromagnetic diffusive effects, thermochemistry and ray-tracing radiative transfer but, the wind description in terms of the Keplerian rotation and mass loss rate is different. \citet{Gressel_2020} determine that the deviation of the winds, caused by magneto-centrifugal mechanism, are super-Keplerian while \citet{Wang_2019} found to be sub-Keplerian. The mass loss in \citet{Gressel_2020} is $\sim10^{-7} M_{\odot}$ yr$^{-1}$ whereas \citet{Wang_2019} is $\sim10^{-8} M_{\odot}$ yr$^{-1}$.

The coexistence between the VSI and photoevaporation processes by stellar photons is of particular interest, especially in regions where we expect steep temperature profile and fast thermal relaxation timescales which few have found (e.g., \citet{2018haex.bookE.138K, Lyra2019, Pfeil2019, Pfeil_2020}) that the VSI might operate down to $\sim$1 AU. The effect of viscous heating in the disk remains relatively small, making it plausible to use passive irradiated disks models to accurately represent the thermal structure \citep{Flock_2019}.  As a first step towards to include photoevaporation processes in our HD simulations, our aim in this first part is to study the convergence, the strength and the region where the VSI operate in the disk, using the full meridional domain. The structure of this paper is as follow: in \S2 we present the methodology in which we describe the disk model and boundary conditions to resolve the VSI using a global 2D isothermal accretion disk configuration. 

In \S3 we present the results with a careful analysis of the kinetic energy and vertical velocity from a global and local perspective. We also provide predictions of the influence that extreme ultraviolet (EUV), far-ultraviolet (FUV) and X-Rays photons can have in the inner parts of the disk. In \S4 we discuss and present a perspective of future work about the implications to planet formation and winds in disk. Finally, \S5 we present our conclusions.

\section{Methods}

We performed our HD simulations using \textsc{PLUTO 4.3}$^{1}$  by \citet{Mignone_2007}. The dynamic of fluids is described by the conservation laws accounting for the divergence of mass density, and momentum density. The HD fundamental equations in our setup are:

\begin{equation}
    \frac{\partial\rho}{\partial t} + \nabla \cdot (\rho\textbf{v}) = 0 
\end{equation}
\begin{equation}    
    \frac{\partial(\rho\textbf{v})}{\partial t} +  \nabla \cdot (\rho \textbf{vv}) + \nabla P  =  - \rho \nabla\Phi
\end{equation} 
\label{stokes_eqs}

where $\rho$ is the mass density, $\textbf{v}$ is the velocity vector, $\rho\textbf{v}$ is the momentum density vector, and $P$ is the gas pressure. We select the isothermal equation of state with $P=c_{\mathrm{s}}^{2}\rho$, where $c_{\mathrm{s}}$ is the isothermal sound speed. The code solve consistently the HD equations in a 2D geometry and considers the system in spherical coordinates ($r,\theta,\phi$) with axisymmetry in the azimuth. The grid cells are set up with a logarithmic increase in the radial domain and a uniform spacing in cos($\theta$) (e.g. \citet{2015MNRAS.447.3512O}) for the meridional domain in order to have a better resolution around the midplane layers. The simulations are scale free. We set the radial domain extending from 0.5 to 5.0 and the vertical domain covering approximately 180$^{\circ}$.  

\footnotetext[1]{\textsc{PLUTO 4.3} is an open source code, to download: \url{http://plutocode.ph.unito.it/download.html}}

\subsection{Disk model and Boundary conditions}

\begin{figure}
\centering
    \includegraphics[width=1.0\linewidth]{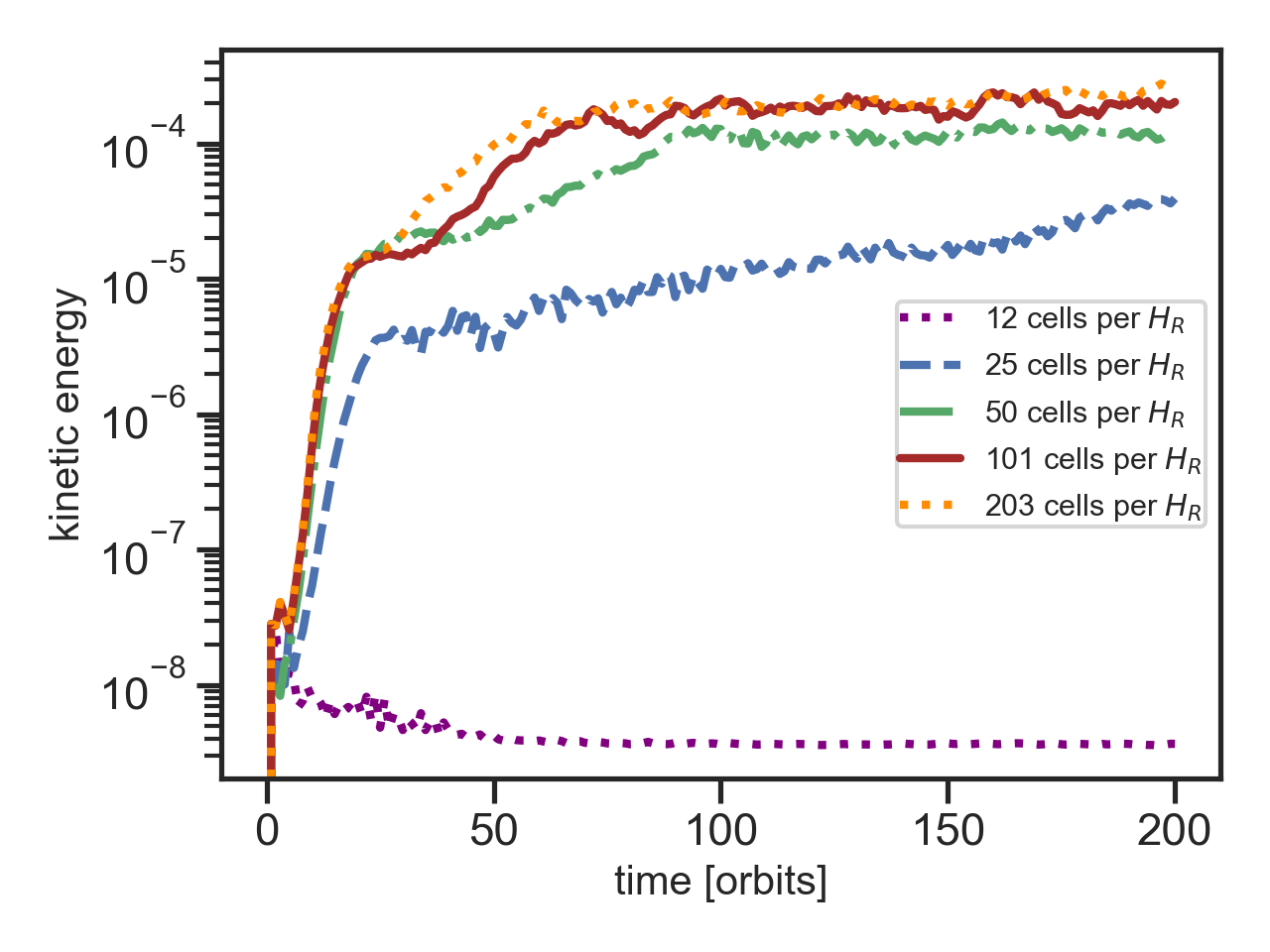}
    \caption{Time evolution of the kinetic energy for different resolutions and space averaged at $R_\mathrm{0}$ = 1 AU. The kinetic energy is normalized with the kinetic energy from pure Keplerian velocity. The different resolutions are color-coded.}
    \label{fig:ke_hr_all}
  
\end{figure}

Protoplanetary disks, rotating with Keplerian angular velocity and vertical stratification are dynamically stable according to Solberg-H\o iland criteria (e.g. \citet{rudiger_2002}). The accretion disk setup in equilibrium in cylindrical coordinates ($R,Z$) \citep{Nelson_2013} are defined:

\begin{equation}
    \label{density}
      \rho(R,Z) = \rho_{\mathrm{0}} \left(\frac{R}{R_{0}} \right)^{p}  \exp{\left (\frac{GM}{c_{\mathrm{s}}^{2}} \left[\frac{1}{\sqrt{R^2 + Z^2} } - \frac{1}{R} \right] \right)} \,, 
\end{equation}

\begin{equation}
      \label{azimuthalvelocity}
      \Omega(R,Z) = \Omega_\mathrm{k} \left[ (p+q) \left(\frac{H_\mathrm{R}}{R} \right)^{2} + (1+q) - \frac{qR}{\sqrt{R^2 + Z^2} } \right]^{1/2}  \,,
\end{equation}

\begin{figure*}[ht]
\centering
\begin{subfigure}[b]{0.33\textwidth}
   \includegraphics[width=\textwidth]{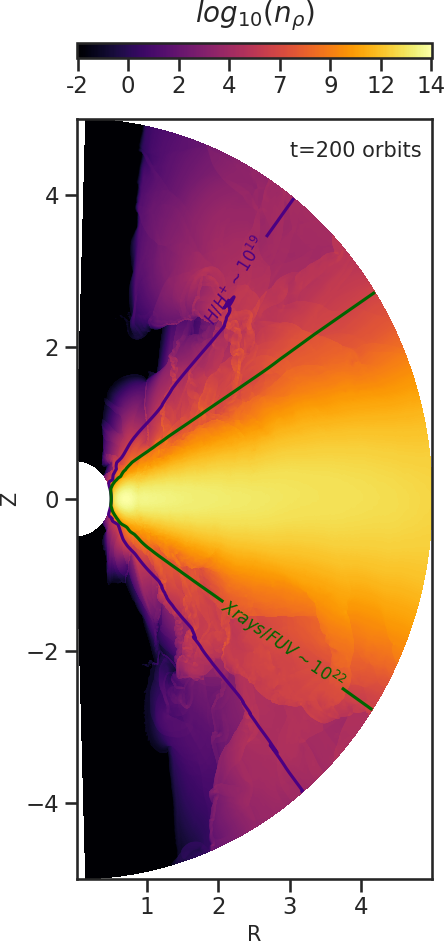}
   \caption{}
\end{subfigure}
\begin{subfigure}[b]{0.326\textwidth}
   \includegraphics[width=\textwidth]{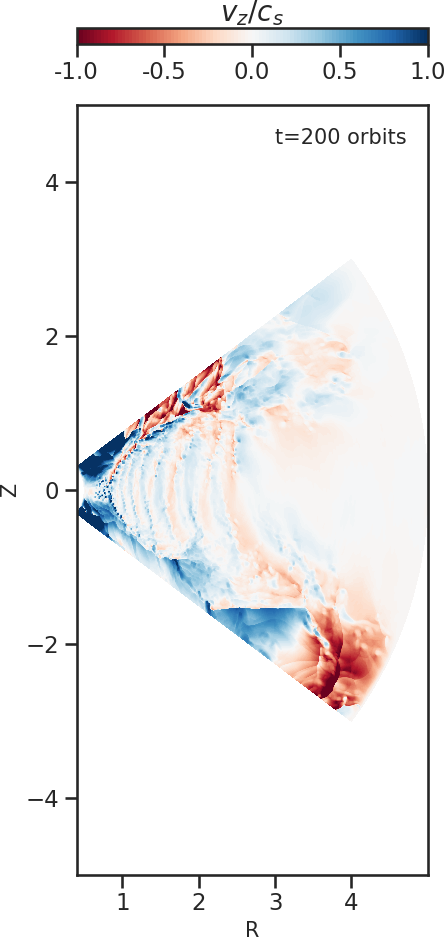}
   \caption{}
\end{subfigure}
\begin{subfigure}[b]{0.333\textwidth}
   \includegraphics[width=\textwidth]{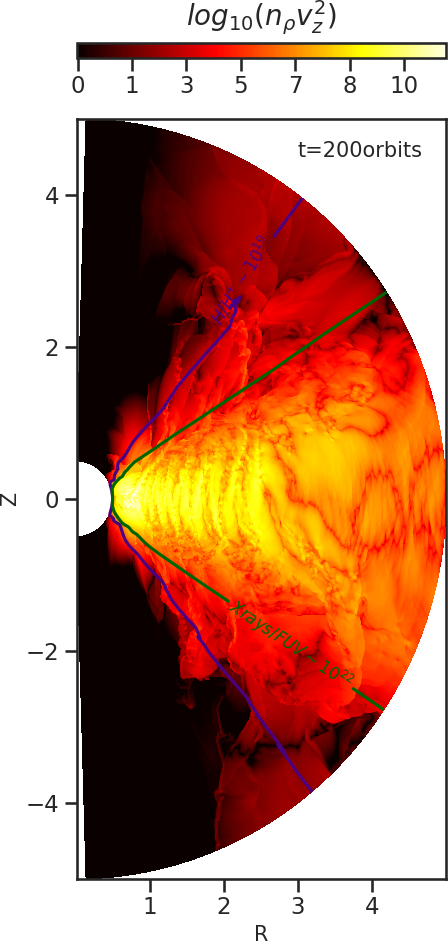}
   \caption{}
\end{subfigure}

\caption{Snapshot of the three main global variables in the simulation. Panel (a) shows the number density in cm$^{-3}$. Panel (b) shows the vertical velocity normalized with respect to the sound speed. For better visibility of the midplane region, we show only an extent of $\pm53^{\circ}$ with respect to the midplane. Panel (c) shows the vertical kinetic energy in g~cm$^{-1}$~s$^{-2}$. Overplotted are contour solid lines where the column density of the EUV-heated region is defined by $\Sigma_{r}$=10$^{19}$ cm$^{-2}$ (in purple), and the column density of the X-rays/FUV-heated boundary layer is defined by $\Sigma_{r}$=10$^{22}$ cm$^{-2}$ (in dark green). The number density of hydrogen nuclei at the H/H$^{+}$ boundary is $\sim$10$^{9}$~cm$^{-3}$ and at the X-rays/FUV-heated layer is $\sim$10$^{12}$~cm$^{-3}$.}

\label{fig:rho_momentum_ke}
\end{figure*}

where $\rho_\mathrm{0}=\frac{\Sigma_{\mathrm{0}}}{\sqrt{2\pi}~H_\mathrm{0}~R_\mathrm{0}}$=4.5$\times$10$^{-10}$g~cm$^{-3}$ = 2.7$\times$10$^{14}$~cm$^{-3}$ is the initial density at the midplane, and $\Sigma_{\mathrm{0}}$ is the initial surface density (see Table\ref{tab:resolution}), $R_\mathrm{0}$ is the reference radius, and $H_\mathrm{0}$ is the reference scale height. $G$ is the gravitational constant, $M$ is the mass of the star. \textit{p} is the power-law fitting exponent of the density profile and $\Omega_\mathrm{k} = \sqrt{GM/R^{3}} $ is the Keplerian frequency. The disk scale height in terms of the radius

\begin{equation}
      \label{eq:hr}
      H_\mathrm{R} = H_\mathrm{0}  \left(\frac{R}{R_\mathrm{0}} \right)^{(q+3)/2}   
      \,
\end{equation}

where the reference scale height ratio is $H_\mathrm{0}$/$R_\mathrm{0}$ = 0.1 (see Table \ref{tab:resolution}). $c_\mathrm{s}$ is related to the temperature through the power-law fitting exponent \textit{q}. The radial profile of $c_\mathrm{s}$ is given based on eq.\ref{eq:hr} so that $c_\mathrm{s}$=$H_\mathrm{R}$/$\Omega_\mathrm{k}$. The scale length of the vertical mode is proportional to $H_\mathrm{0}$.

In order to compute the fluxes in each cell, we employ second-order piece-wise \textit{linear} spatial reconstruction. This reconstruction method is appropriate for both, uniform and non-uniform grid spacing based on the approach from \citet{Mignone_2014}. We use Harten, Lax, Van Leer (HLL) Riemann solver. For the time integration we adopt second-order Runge-Kutta and we set the Courant-Friedrichs-Lewy (CFL) number to 0.25. Table.\ref{tab:resolution} summarize the model parameter and the different runs for each resolution. For the 101 and 203 cells per $H_\mathrm{R}$, the $\theta$ inner limit is shifted slightly inwards due to high computational precision issue when increasing the resolution at the $\theta$ boundary. We investigated the PPM3 spatial reconstruction using RK3 for the time integration for all resolution cases. We found that the PPM3 could resolve the VSI for even lower resolutions, however, our PPM3 configuration is currently not yet stable. Therefore, we only show our analysis and discussion for the VSI using the \textit{linear} reconstruction method for this paper.
 
\begin{figure*}[]
\centering

\includegraphics[width=1.0\linewidth]{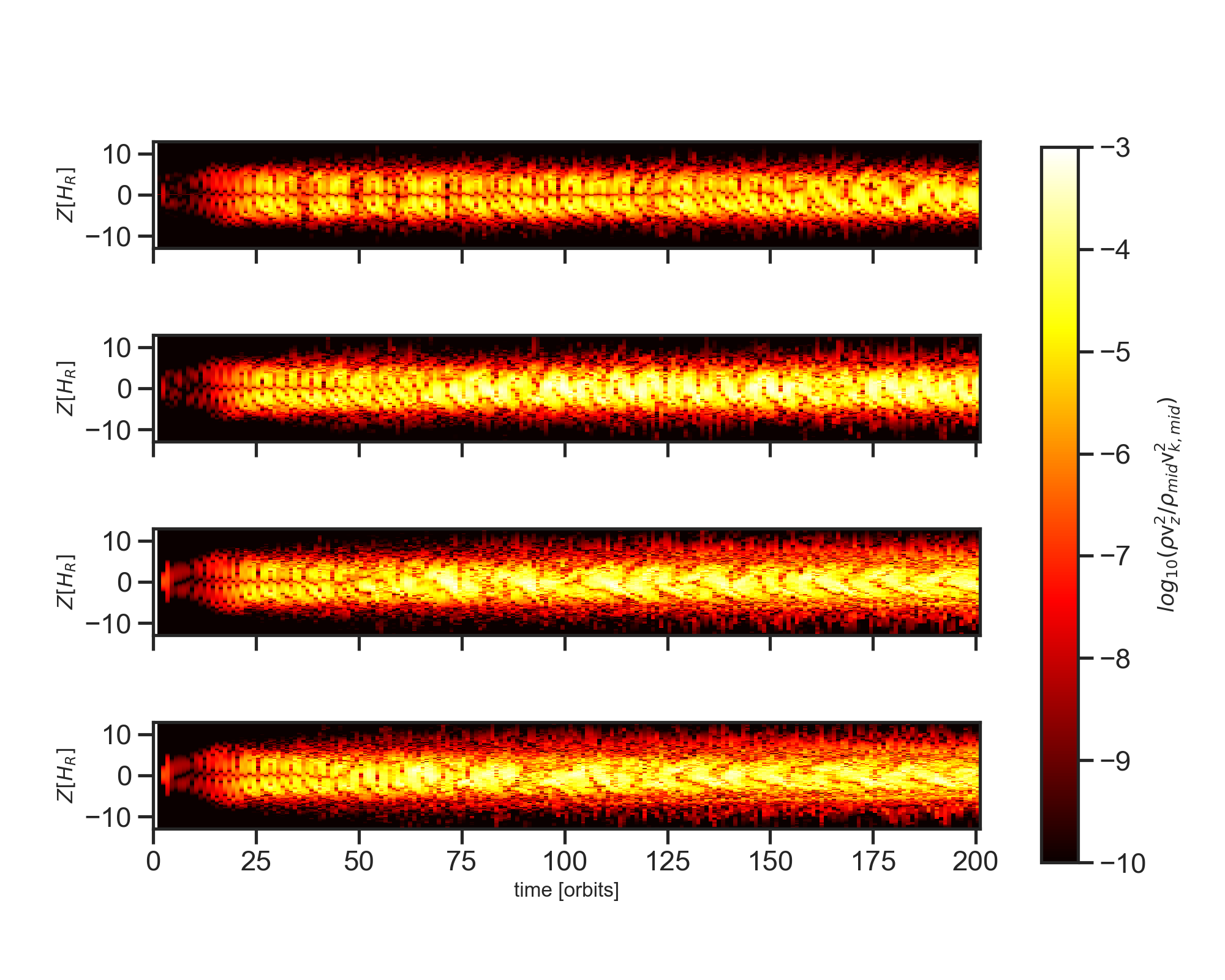}
\caption{Time-evolution of the local kinetic energy at $R_\mathrm{0} = 1$ over height for different resolutions (from top to bottom: 25, 50, 101, 203 cells per scale height). The local kinetic energy is normalized with that to the global kinetic energy (see eq.\ref{eq:ke_norm}).}
\label{fig:local_ke_scale_height}
\end{figure*}

\subsection{EUV,FUV and X-Rays column density}
\label{sec:raytracing}

The inner parts of the disk are highly influenced by the stellar radiation potentially ionizing the upper layers of the disk. As a first step to implement photoheating processes caused by stellar EUV, FUV, and X-rays photons, we calculate the radial column density locally to predict at what scale height the EUV- and X-rays/FUV-heated layers are most dominant. The radial column density is calculated by integrating the hydrogen nuclei column density along the radial line of sight

\begin{equation}
      \Sigma_{r} (r) =  \int_{R_\mathrm{*}}^{r} n_\mathrm{\rho} dr,,
\end{equation}

where $R_{*}$ is the star radius, $n_\mathrm{\rho}$ is the number density of hydrogen nuclei. For this first part, since we do not consider detailed chemistry or frequency of UVs, we determine the boundaries in terms of $\Sigma_{r}$ by post-processing. We focus on the definition of the radial column density to locate the boundary of the ionized region (EUV-heated region) and the neutral layer (X-rays/FUV-heated layer) which is empirically determined from the simulations of \citet{Nakatani_2018a, Nakatani_2018b}. We select the boundary between the ionized or EUV-heated region namely H/H$^{+}$ boundary, to be located at $\Sigma_{r}$=10$^{19}$ cm$^{-2}$, assuming the EUV stellar luminosity is $\sim10^{30}$~erg~s$^{-1}$. The EUV-heated region typically has temperatures of orders of magnitude between 10$^{3}$ to 10$^{4}$ K and the hydrogen is fully ionized here. The boundary of the X-rays/FUV-heated layer is defined at $\Sigma_{r}$=10$^{22}$ cm$^{-2}$ where the gas is heated to 10$^{3}$ K and it dominates in the neutral layers. Though, soft X-Rays are considered to be an important component for photoevaporation, the locations of FUV- and X-Rays-heated layers are nearly identical $\Sigma_{r} \leq 10^{21-22}$ cm$^{-2}$ \citep{Gorti_2008, Nakatani_2018b}. Therefore, setting the boundary of the neutral photo-heated layer at $\Sigma_{r} \leq 10^{21-22}$ cm$^{-2}$ is valid even in cases where X-Rays effects are included. Note that we neglect scattered light from our ray-tracing. For more information about the effects of stellar photons and the launch of winds, see \S\ref{sec:frequency-integrated_cross_section} and \S\ref{sec:electron_abundance}.

\begin{table}[htb]
   \caption{Grid set up and initial parameters for the different runs.} 
   \label{tab:resolution}
   \small 
   \centering 
   \begin{tabular}{lccr} 
   \noalign{\smallskip} \hline \hline
   \noalign{\smallskip} \textbf{$N_\mathrm{r} \times N_\mathrm{\theta} \times N_\mathrm{\phi}$} & \textbf{$\Delta r~[AU]$} & \textbf{$\Delta \theta~[AU]$} \\  
   \hline
   256 $\times$ 256  $\times$ 1 & 0.5:5 & 0.001:3.1405 \\ 
   512 $\times$ 512 $\times$ 1  & 0.5:5 & 0.001:3.1405 \\ 
   1024 $\times$ 1024 $\times$ 1  & 0.5:5 & 0.001:3.1405 \\ 
   2048 $\times$ 2048 $\times$ 1 & 0.5:5 & 0.01:3.1315 \\ 
   4096 $\times$ 4096 $\times$ 1  & 0.5:5 & 0.02:3.1216 \\ 
   \hline
\multicolumn{3}{c}{Initial parameters} \\
   \hline

\textit{p} & -1.5 \\ 
\textit{q} & -1.0 \\
\textit{$R_\mathrm{0}$} & 1.0 \\
\textit{$H_\mathrm{0}$} & 0.1 \\
$\Sigma_\mathrm{0}$ [g~cm$^{-2}$] & 1700\\
\noalign{\smallskip} \hline \noalign{\smallskip}

\end{tabular}

\end{table}

\section{Results}

\begin{figure}
\centering
    \includegraphics[width=1.0\linewidth]{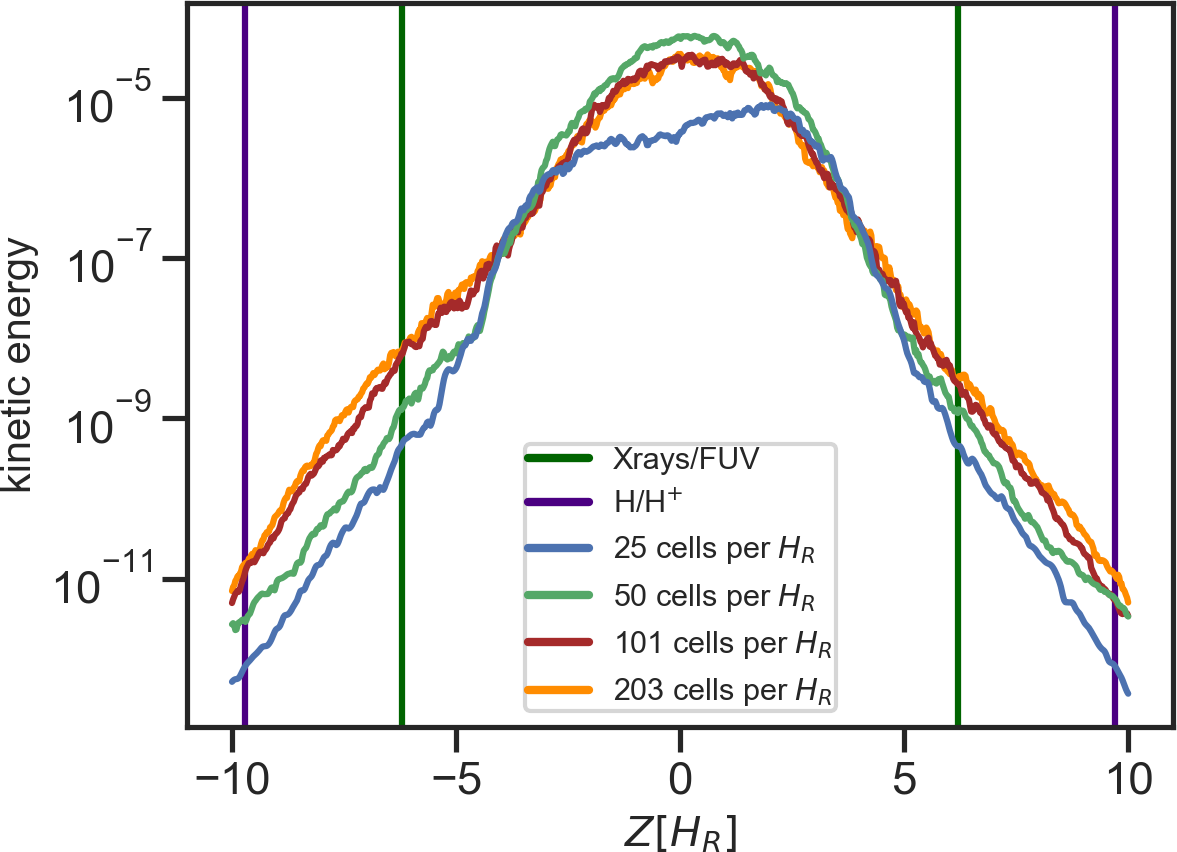}
    \caption{Averaged profile of the local kinetic energy normalized with respect to the kinetic energy in the midplane for the 25, 50, 101, and 203 cells per $H_\mathrm{R}$  at $R_\mathrm{0}$ = 1 AU. The vertical colored lines represent the $H_\mathrm{R}=6.2$, where the X-rays/FUV-heated boundary layer (in dark green) is, and the $H_\mathrm{R}=9.7$, where EUV-heated region (in purple) is located (same as in Fig.\ref{fig:rho_momentum_ke}). The 101 and 203 cells per $H_\mathrm{R}$ are in good agreement. The density floor is located $H_\mathrm{R}$=13, therefore, not influencing our result.}
    \label{fig:ke_midplane_line_plot}
    
\end{figure}

\subsection{Dynamical evolution}
\label{sec:global_ke}
In order to investigate the kinematics we start by looking at the time evolution of the kinetic energy. Fig.\ref{fig:ke_hr_all} shows the averaged kinetic energy at different grid cell resolutions. Our prescription of the global kinetic energy follows

\begin{equation}
      E_{kin} = \frac{1}{2} \int_{v} \rho v_{Z}^{2} dV
        \label{eq:k.e.}
\end{equation}

where $v_{Z}$= $v_{r}$$\cos(\theta)$ - $v_{\theta}$$\sin(\theta)$ is the vertical component of the velocity and $E_{kin}$ is normalized with respect to

\begin{equation}
      E_{norm} = \frac{1}{2} \int_{v} \rho v_{k}^{2} dV
      \label{eq:ke_norm}
\end{equation}

where $v_{k}$ is the Keplerian velocity. Due to our large global domain, it becomes necessary to transform our spherical velocity components to cylindrical when describing the vertical kinetic energy. For the resolution analysis we simulate a large parameter space, starting from 12 cells per scale height until 203 cells per scale height. At the lowest resolution, we observed that the VSI is not resolved as the kinetic energy is not growing (see Fig.\ref{fig:ke_hr_all}). We observe that the initial perturbations are damped and the kinetic energy is decreasing leading to a fully laminar disk. For the resolutions higher than 12 cells per scale height, the growth seen at $\geq$25 orbits is characteristic of the initial increase phase of the kinetic energy as the disk become unstable to the VSI, similar as found by \citet{Nelson_2013}. The 25 cells per scale height case is often used in the literature as a standard resolution to perform global simulations. However, we emphasize that this resolution does not fully solve the VSI in accretion disks. It reaches an energy saturation slightly less than 10$^{-5}$ and after that, it starts to increase monotonically without reaching to steady state at least within the first 200 orbits. Because it is characteristic for all resolutions, the first energy saturation $\sim$10$^{-5}$ is a consequence of the linear phase of the instability spreading throughout the system. The second phase of increasing kinetic energy might be related to the slower growth of the body modes that start to become visible after 25 orbits, when the fast growing finger modes have already saturated. During this second phase, the body modes are growing in the disk until they reach saturation after around 90 orbits. Our results show that at a resolution of about 100 cells per scale height the kinetic energy level converge. For even higher resolutions, which are computationally very expensive, we predict to have the same convergence trend. 

Comparing our total kinetic energy, it follows the same trend as the perturbed kinetic energy from \citet{Nelson_2013}. The resolution for the VSI used by \citet{Cui_2020} lies in between 96 and 108 cells per scale height which is in good agreement to our resolution study as well. Fig.\ref{fig:rho_momentum_ke} shows a snapshot of our three main physical quantities in our analysis: the number density, the vertical velocity, and the kinetic energy in physical units. Overplotted are the contour lines in which the EUV-heated region and the X-rays/FUV-heated boundary layer are defined based on our definition of $\Sigma_{r}$ (\S\ref{sec:raytracing}). 

The density (Fig.\ref{fig:rho_momentum_ke} panel (a)) has a smooth profile until reaching the upper layers. Our density floor starts to become important at $H_\mathrm{R}$=13 which lies in a region where we do not expect VSI activity.
As expected, the FUV photons penetrate deeper than EUV photons within the first 5 AU (Fig.\ref{fig:rho_momentum_ke}). The X-rays/FUV-heated layer and EUV-heated region reach a $H_\mathrm{R}=6.2$ and $H_\mathrm{R}=9.7$, respectively. At this photoevaporation region we predict that HI regions becomes optically thick against EUV with a $\Sigma_{r}$=10$^{19}$ cm$^{-2}$ to $\Sigma_{r}$=10$^{22}$ cm$^{-2}$ (Fig.\ref{fig:column_density}). 

The profile of the vertical velocity (Fig.\ref{fig:rho_momentum_ke} panel (b)) shows clearly the upward motion shaded in blue and the downward motion shaded in red. In this region, the vertical velocity reach a fraction of the sound speed.

The kinetic energy profile (Fig.\ref{fig:rho_momentum_ke} panel (c)) shows the highest concentration present as corrugated-shape like structures resembling the body modes in the disk. In the next section, we investigate this in details in a more local perspective.

\begin{figure}
\centering
    \includegraphics[width=1.0\linewidth]{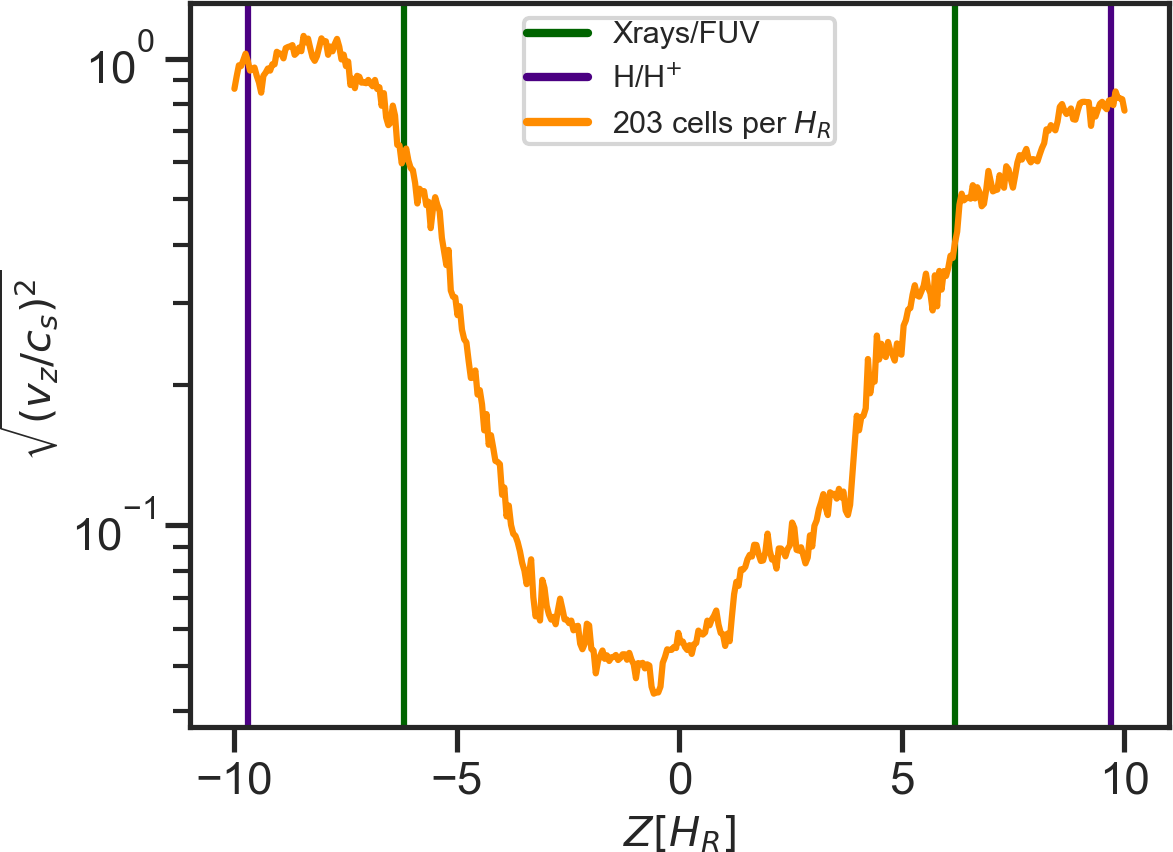}
    \caption{Averaged profile of the vertical component of the velocity normalized with respect to the sound speed for the 203 cells per $H_\mathrm{R}$. The vertical colored lines are the same as in Fig.\ref{fig:ke_midplane_line_plot}. At $H_\mathrm{R}=$6.2, the vertical velocity reach to 0.3$c_\mathrm{s}$. We expect the hot ionized region to start beyond $H_\mathrm{R}>6.2$.}
    \label{fig:vz_cs}

\end{figure}

\subsection{Local kinetic energy evolution over height and time}
\label{sec:local_ke}

We show the time-evolution of the kinetic energy over $H_\mathrm{R}$ at $R_\mathrm{0}$ in Fig.\ref{fig:local_ke_scale_height}. We transform the velocity vectors to cylindrical and, in order to see the local kinetic energy in a more continuous spatial structure, we then use a cubic method to interpolate the data on a new grid space with a $Z$ direction betwen -1 and 1 AU at $R_\mathrm{0}$. Fig.\ref{fig:local_ke_scale_height} shows that the kinetic energy is growing first the finger modes in the upper layers of the disk, which is expected. At the midplane, there is no VSI activity since the vertical shear vanishes right at the midplane. Once the body modes grow, the large scale velocity perturbations are threatening the upper and lower hemisphere. At this stage also the kinetic energy at the midplane grows. 

Two important features are depicted from all resolution cases in Fig.\ref{fig:local_ke_scale_height}. The first one is the presence of vertical lanes of kinetic energy initially. We refer these vertical kinetic energy lanes as finger modes and they are consistent with the linear energy increase phases from Fig.\ref{fig:ke_hr_all}. The second feature corresponds to the appearance of the body modes. As seen in Fig.\ref{fig:local_ke_scale_height}, the growth rate of the body modes increases with resolution. For the 203 cells per scale height, the body modes saturate at around $\sim$70 orbits. At $H_\mathrm{R}$=0 (midplane), the kinetic energy is very low but even more detailed analysis in the midplane is performed in the last paragraph.

Fig.\ref{fig:ke_midplane_line_plot} shows the time averaged vertical profile of the kinetic energy. The vertical profile follow a similar trend as the density profile. Only for the lowest resolution case, we observe a plateau feature, which is due to the insufficient resolution to solve for the VSI. The VSI seems to operate in the full vertical extent. We expect a physical boundary for the VSI to be the high temperature region which is caused by the FUV radiation. We also expect this position to be where the X-rays/FUV-heated boundary layer is located at $H_\mathrm{R}$=6.2 (see Fig.\ref{fig:ke_midplane_line_plot} \& Fig.\ref{fig:vz_cs}). Fig.\ref{fig:vz_cs} shows that the time averaged vertical velocity increases to levels of several tens of percent at the expect wind base at $H_\mathrm{R}$ around 6.2. In the midplane layers, the time averaged vertical motions remain small, on a level around 1$\%$.

\begin{figure}
\centering
    \includegraphics[width=1.0\linewidth]{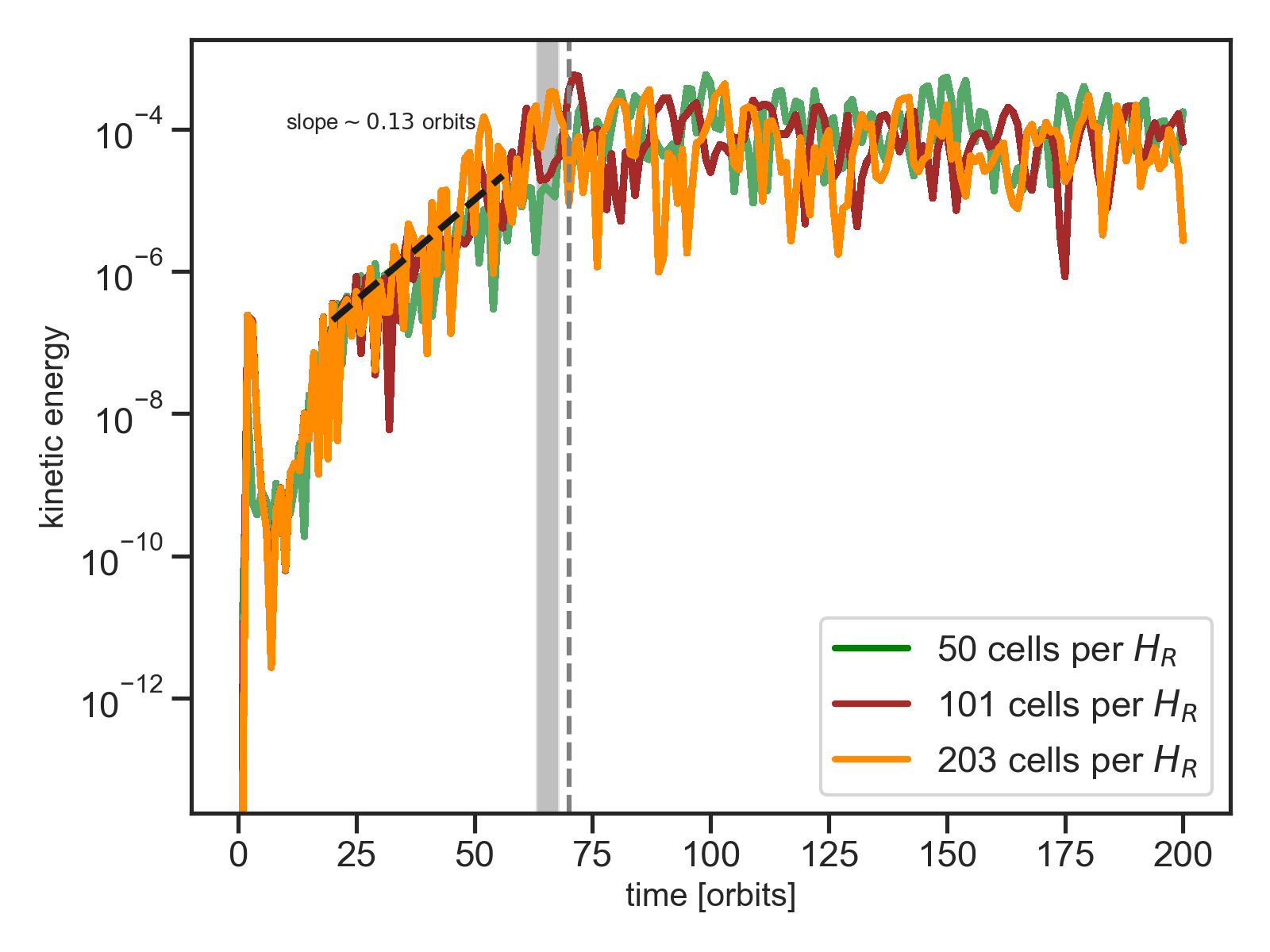}
    \caption{Time evolution of the local kinetic energy in the midplane for the 50, 101, and 203 cells per scale height resolution case at $R_\mathrm{0}$. The black dashed line represent the slope of the linear growth rate phase of the local energy in the midplane. The vertical gray lane represent about the time where the 101 and 203 cells per scale height resolutions reaches saturation and the vertical dashed one represent about the time where the 50 cells per scale height converges onto stability.}
    \label{fig:ke_midplane}
\end{figure}

Fig.\ref{fig:ke_midplane} shows the time-evolution of the local kinetic energy at the midplane at $R_\mathrm{0}$. The local kinetic energy at the midplane replicates the same two growth regimes as the averaged global kinetic energy seen in Fig.\ref{fig:ke_hr_all}. The fast growth from the finger modes that can even be seen at the vertical velocity at the midplane, and the slow growth of the body modes. During the growth phase of the generation of the body modes, the local kinetic energy in the midplane in all three resolutions follow a linear growth rate of $\sim$0.13~orbits. This growth phase persists within the first $\sim$70 orbits and it corresponds to the slow growth rate of the body modes until reaching a saturation level of $\sim$10$^{-4}$ and it is 2 times slower compared to the general growth rate found by \citet{Nelson_2013}. The 203 and 101 cells per scale height generate the body modes faster than the 50 cells per scale height, which is the same tendency as found in Fig.\ref{fig:local_ke_scale_height}. We conclude that we need higher than 50 cells per scale height to solve the VSI in accretion disks.

\subsection{Small scale vortices}
\label{sec:small_vortices}

\begin{figure*}[]
\centering
\begin{subfigure}{0.59\textwidth}
   \includegraphics[width=\textwidth]{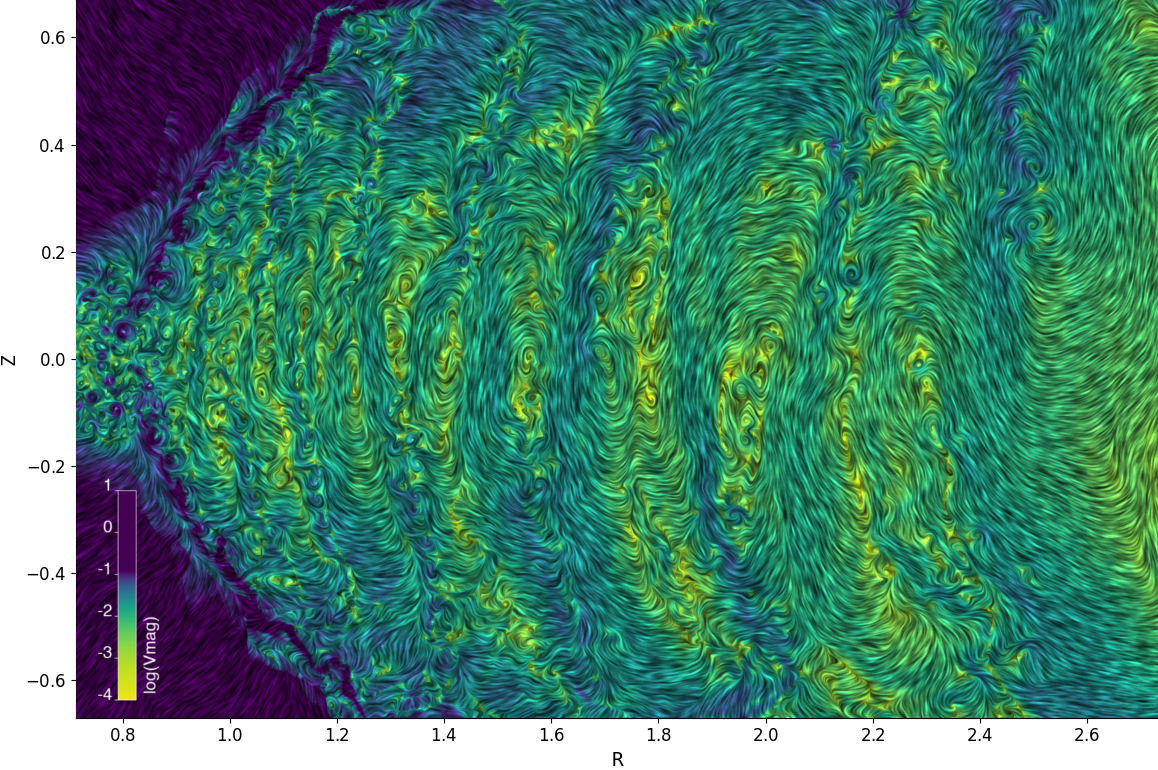}
   \caption{}
\end{subfigure}

\begin{subfigure}{0.59\textwidth}
   \includegraphics[width=\textwidth]{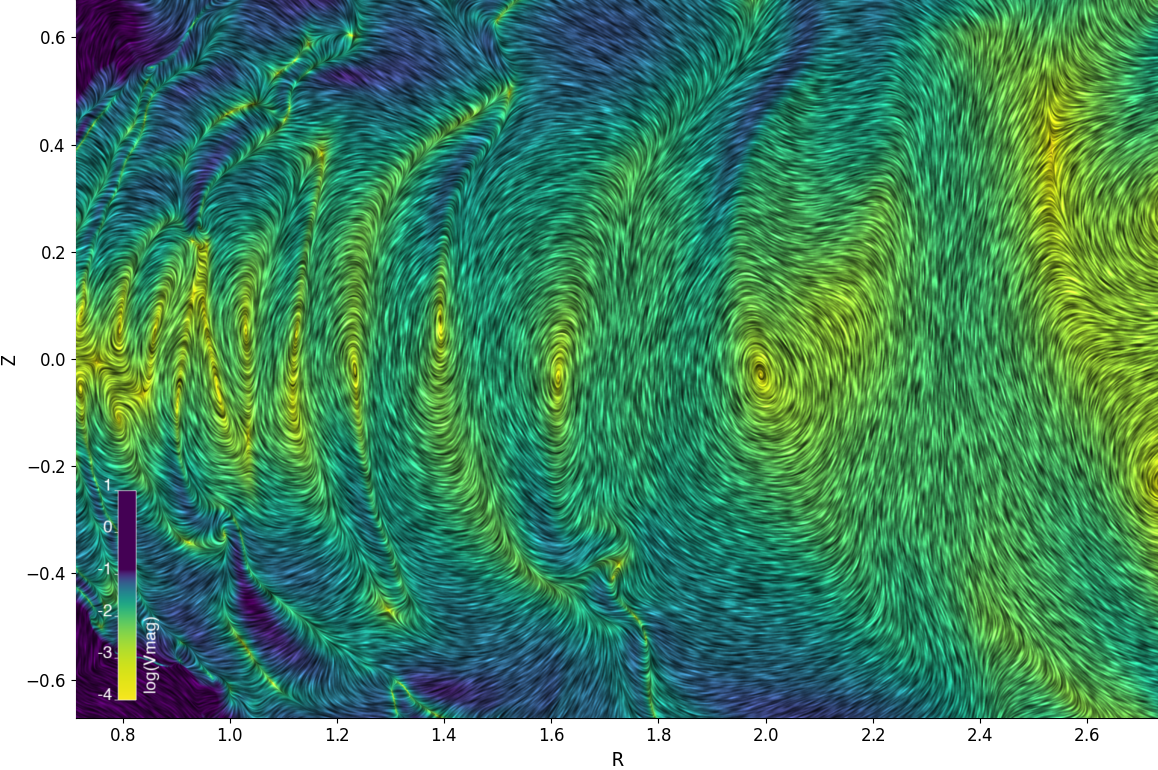}
   \caption{}
\end{subfigure}

\caption{Snapshot of $v_\mathrm{mag}$ for the 203 cells per scale height (Panel (a)) compared to the lowest resolution of 25 (Panel (b)) cells per scale height at 200 orbits. The color contour plot shows $v_\mathrm{mag}$ and the streamlines show the velocity flow pattern when applying the LIC method. The $v_\mathrm{mag}$ range (from 0 to 0.2 in code units) is narrowed down to highlight the low velocity perturbations close to the midplane associated to the VSI.}
\label{fig:LIC}
\end{figure*}

In Fig.\ref{fig:LIC}, we show the result of the Line Integral Convolution (LIC) method (e.g. \citet{Cabral_2003}) to visualize the velocity magnitude vector field associated to velocity perturbations from the VSI, for the lowest and highest resolutions. The velocity magnitude includes the velocity field in every component and follows the ordinary prescription of $v_\mathrm{mag}=\sqrt{v_\mathrm{Z}^{2}+v_\mathrm{R}^2}$ where $v_\mathrm{R}$=$v_{r}$$\sin(\theta)$ + $v_{\theta}$$\cos(\theta)$. From the LIC method, we notice small scale vortices in both resolutions cases, however, these small scale vortices are not featured in the lowest resolution case.

The possibility of the presence of vortices formed by VSI was first found by \citet{Richard_2016} in the $r - \phi$ plane. These vortices can be long-lived locally around 500 orbits \citep{Manger2018} with an $H_\mathrm{R}$ = 0.1 and with an azimuthal and radial extension of about 40 and 4 au \citep{Flock_2020}, respectively. In our case the vortices appear in the $r - Z$ plane, mostly inside the region between two large scale upward and downward motions. Moreover, these small scale vortices happen in an $H_\mathrm{R} < 5$, therefore, being more important for dust evolution.

\section{Discussion of implications to planet formation and winds in disk}

The VSI is a large scale global instability that potentially plays an important role for the evolution and dynamics of protoplanetary disks. In the midplane, the dust settling overcomes the VSI, and as the dust concentrates and grows to mm size and greater, these particles settle against VSI and eventually undergoes SI to favor planet formation \citep{Lin2019}.  However, the VSI in combination with the SI leads to radial dust concentration in long-lived accumulations which are significantly denser areas than those formed by the SI alone \citep{2020A&A...635A.190S}. 

We observe symmetric features, which have not been observed before, after 100 orbits in the time evolution of the kinetic energy (Fig.\ref{fig:local_ke_scale_height}). Furthermore, we refer to the LIC plots (Fig.\ref{fig:LIC}) that shows the presence of small vortices along the midplane, where grains can be concentrated . We encourage further analysis of this manner and its effect on planet formation in the future.

\subsection{Importance of stellar photons layers and the launch of winds}
\label{sec:importance_photons_in_winds}

The VSI can maintain small particles ($<100~\mu$m) lofted \citep{Lin2019} until eventually becoming ionized by FUV photons at the disk surface and contributing to heating in the atmosphere. Both, the photoionization effects and the high temperature in this region, makes the surface of the disk and atmosphere an optimal scenario for photoevaporation processes, resulting in the launch of winds. In the disk atmosphere, the MHD winds are likely favorable \citep{Cui_2020} at less than 5 AU. \citet{Cui_2020} showed that in fact VSI and MHD winds can coexist with magnetized winds the main driver of angular momentum transport. However, this would rather depend on 1) the so-called wind base which depend mainly on the position of the EUV and FUV layers and 2) a realistic thermal profile. By considering the initial flux and the extinction along the radial line of sight we can determine the location and extension of the wind in the future work. The detailed role of the VSI for the mechanism of wind launching has to be investigated, especially due to the fact that the magnetic coupling to the gas remains low in the FUV shielded region $H_\mathrm{R}<6.2$. Recent improvements, towards including a more realistic scenarios, by \citet{Wang_2019} and \citet{Gressel_2020} have included thermochemistry and irradiation in their global magnetohydrodynamical simulations.

\subsubsection{Frequency-integrated cross-sections}
\label{sec:frequency-integrated_cross_section}

Perhaps the most robust way to define the influence of EUV and X-rays/FUV photons layers is by taking into consideration the frequency-dependent cross-sections for each photon respectively. The radial optical depth of each photon depends on the frequency range that is used in the integration of the ionization rates and the associated heating rates. This is very important to consider when characterizing the ionized region in the disk atmosphere, that is influenced by EUV photons. The atomic hydrogen abundance in this ionized region is negligible up in the ionization front, where we expect the radial optical depth of the EUV equal to 1 \citep{Hollenbach_1994, Tanaka_2013, Nakatani_2018a, Nakatani_2018b}. This boundary between the ionized region and the neutral layer is rather located at an atomic hydrogen column density of $10^{18}$ cm$^{-2}$. This atomic hydrogen column density can be translated into a hydrogen column density of $10^{19-20}$ cm$^{-2}$ depending on the stellar EUV luminosity and, therefore, used here to define the EUV-heated region.

Given the fact that FUV photons can promote heating via photoelectric heating \citep{Bakes_1994} in the disk atmosphere, such photons are attenuated by dust once the hydrogen column density is in between 10$^{21}$ cm$^{-2}$ to 10$^{22}$ cm$^{-2}$ \citep{Gorti_2009, Nakatani_2018a}. When considering the radial optical depth of the FUV, the FUV opacity evaluated for graphite is about $\sim$10$^{4}$ cm$^{2}$~g$^{-1}$ that is in the wavelength regime between 0.2 to 0.09 $\mu$m (6 eV $< h\nu <$ 13.6 eV). Assuming a dust-to-gas mass ratio of 0.01 this gives a total FUV opacity of 100 cm$^{2}$~g$^{-1}$. Though, it is more likely that the dust-to-gas mass ratio is rather smaller than 0.01 in the upper layers, perhaps a more nominal range could be somewhat from 0.001 to 0.0001, shifting the radial optical depth of the FUV equal to 1 closer to the midplane. This FUV opacity prescription is consistent with the dust population used by \citet{Mario_2016} to describe the physical properties of the inner rim.  In realistic disks scenarios, which we consider for the second part of this project, the EUV and FUV flux relies on the wavelength-dependent initial flux we assume and the extinction along the radial line of sight at each location in the disk. Meanwhile, we expect the EUV to provide an important effect for the extension of the wind in the disk atmosphere.

\subsubsection{Electron density}
\label{sec:electron_abundance}

The temperature and electron number density can be inferred from thermally excited forbidden lines (e.g. [OI] 5577/6300 line ratio) in regions where high temperatures (> 5000 K) lead to collisional excitations of electrons \citep{2Simon_2016}. Including or not the effects of X-Rays, the electron relative abundance between the EUV-heated and the X-rays/FUV-heated layer is still $\sim$ $10^{-4}$ that is coming from atomic carbon ionization by FUV. The electron density in this region can be as high as $\sim$ $10^{4}$-$10^{5}$ cm$^{-3}$ at $\sim$1 AU \citep{Nakatani_2018a, Nakatani_2018b}. We expect that at $<$ 1 AU, the X-Rays, EUV, and FUV fluxes are higher, therefore, photons penetrate deeper in the disk and the electron density can be higher. On the other hand, the farther from the star, the electron abundance decreases and so the electron density.

\section{Conclusions}

We have performed 2.5D (axisymmetric) hydrodynamical simulations for an isothermal accretion disk with different resolutions, up to 203 cells per scale height to fully resolve for the VSI. Resolution study for the VSI is important to assess its effect in different regions of the disk. We determine that the standard resolution of 50 cells per scale height to be the lower limit to resolve the VSI. This determination is based on the total kinetic energy saturation level. Our key points can be summarize as

   \begin{enumerate}
     \item We found the VSI to operate throughout the entire disk until reaching our density floor 0.003~cm$^{-3}$ at around $H_\mathrm{R}=13$ at $R_\mathrm{0}$=1 for resolutions $\geq$ 50 cells per scale height. We expect turbulent scenarios where the X-rays/FUV-heated boundary layer at $H_\mathrm{R}=6.2$ is located, whereas, the EUV-heated region reach a $H_\mathrm{R}=9.7$. Nevertheless, we encourage the use for high resolution ($\geq$ 50 cells per scale height) to properly conduct analysis of VSI. 
     \item We found at a resolution of 25 cells per $H_\mathrm{R}$ the appearance of a plateau around the midplane when looking at the local kinetic energy over height. This could serve as a diagnostic to check whether or not the VSI is fully resolved.
     \item We observe a clear dependence on the growth rate of the body modes with resolution. For the highest resolution the growth of the body modes saturates after 70 orbits.
     \item We diagnose the presence of small scale vortices between the large scale motion of the VSI when performing the LIC method. These small scale vortices reach a $H_\mathrm{R}<5.0$, therefore, being more important for dust evolution. The vertical kinetic energy shows a more symmetric oscillation for the upper and lower hemisphere for the highest resolution. In the future, we will fully investigate the dynamical influence of the heated layers on the VSI and the disk evolution. A resolution of 50 cells per scale $H_\mathrm{R}$ seems to capture the main characteristics of the VSI and so suitable for future studies. 
   \end{enumerate}
   
\begin{acknowledgements}
      This work is supported by the \emph{European Research Council (ERC)} project under the European Union's Horizon 2020 research and innovation programme number 757957. The LIC images were generated by Thomas M{\"u}ller from the Haus dier Astronmomie. We would like to thank the matplotlib team \citep{Hunter_2007} for the good quality tool in order to better visualize the data. 
\end{acknowledgements}
\vfill

\appendix

\section{Radial column density}
\label{sec:radial_column_density}

\begin{figure}[hbt!]
\centering
    \includegraphics[width=0.57\linewidth]{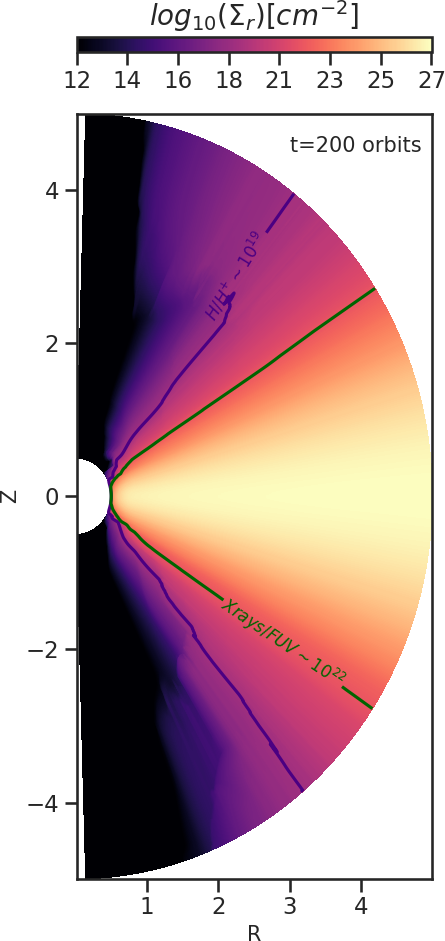}  
    \caption{Radial column density in cm$^{-2}$. Overplotted are solid lines where: $\Sigma_{r}$=10$^{19}$ cm$^{-2}$ represent the EUV-heated region (in purple), and $\Sigma_{r}$=10$^{22}$ cm$^{-2}$ represent the X-rays/FUV-heated boundary layer (in dark green).}
    \label{fig:column_density}
\end{figure}

We conduct a surface density analysis for the determination of the EUV-heated region and X-rays/FUV-heated boundary layer (see \S\ref{sec:raytracing}). Fig.\ref{fig:column_density} shows the contour plot of the radial column density and overplotted are the two regions of interest. 

\bibliographystyle{aa}
\bibliography{VSI} 

\begin{thebibliography}{51}
\expandafter\ifx\csname natexlab\endcsname\relax\def\natexlab#1{#1}\fi

\bibitem[{{Alexander} {et~al.}(2014){Alexander}, {Pascucci}, {Andrews},
  {Armitage}, \& {Cieza}}]{Alexander_2014}
{Alexander}, R., {Pascucci}, I., {Andrews}, S., {Armitage}, P., \& {Cieza}, L.
  2014, in Protostars and Planets VI, ed. H.~{Beuther}, R.~S. {Klessen}, C.~P.
  {Dullemond}, \& T.~{Henning}, 475

\bibitem[{{Alexander} {et~al.}(2006{\natexlab{a}}){Alexander}, {Clarke}, \&
  {Pringle}}]{Alexander_2006I}
{Alexander}, R.~D., {Clarke}, C.~J., \& {Pringle}, J.~E. 2006{\natexlab{a}},
  \mnras, 369, 216

\bibitem[{{Alexander} {et~al.}(2006{\natexlab{b}}){Alexander}, {Clarke}, \&
  {Pringle}}]{Alexander_2006}
{Alexander}, R.~D., {Clarke}, C.~J., \& {Pringle}, J.~E. 2006{\natexlab{b}},
  \mnras, 369, 229

\bibitem[{Arlt \& Urpin(2004)}]{arlt2004simulations}
Arlt, R. \& Urpin, V. 2004, Astronomy \& Astrophysics, 426, 755

\bibitem[{{Bakes} \& {Tielens}(1994)}]{Bakes_1994}
{Bakes}, E.~L.~O. \& {Tielens}, A.~G.~G.~M. 1994, \apj, 427, 822

\bibitem[{{Ballabio} {et~al.}(2020){Ballabio}, {Alexander}, \&
  {Clarke}}]{Ballabio_2020}
{Ballabio}, G., {Alexander}, R.~D., \& {Clarke}, C.~J. 2020, \mnras, 496, 2932

\bibitem[{{Banzatti} {et~al.}(2019){Banzatti}, {Pascucci}, {Edwards}, {Fang},
  {Gorti}, \& {Flock}}]{Banzatti2019}
{Banzatti}, A., {Pascucci}, I., {Edwards}, S., {et~al.} 2019, \apj, 870, 76

\bibitem[{{{Barker, A.J. and Latter, H.N.}}(2015)}]{Barker_2015}
{{Barker, A.J. and Latter, H.N.}} 2015, \mnras, 450, 21

\bibitem[{Cabral \& Leedom(1993)}]{Cabral_2003}
Cabral, B. \& Leedom, L.~C. 1993, in Proceedings of the 20th Annual Conference
  on Computer Graphics and Interactive Techniques, SIGGRAPH ’93 (New York,
  NY, USA: Association for Computing Machinery), 263–270

\bibitem[{{Clarke} {et~al.}(2001){Clarke}, {Gendrin}, \&
  {Sotomayor}}]{Clarke_2001}
{Clarke}, C.~J., {Gendrin}, A., \& {Sotomayor}, M. 2001, \mnras, 328, 485

\bibitem[{Cui \& Bai(2020)}]{Cui_2020}
Cui, C. \& Bai, X.-N. 2020, The Astrophysical Journal, 891, 30

\bibitem[{{Dzyurkevich} {et~al.}(2013){Dzyurkevich}, {Turner}, {Henning}, \&
  {Kley}}]{Dzyurkevich_2013}
{Dzyurkevich}, N., {Turner}, N.~J., {Henning}, T., \& {Kley}, W. 2013, \apj,
  765, 114

\bibitem[{{Ercolano} {et~al.}(2008){Ercolano}, {Drake}, {Raymond}, \&
  {Clarke}}]{Ercolano_2008}
{Ercolano}, B., {Drake}, J.~J., {Raymond}, J.~C., \& {Clarke}, C.~C. 2008,
  \apj, 688, 398

\bibitem[{{Ercolano} \& {Pascucci}(2017)}]{Ercolano_2017}
{Ercolano}, B. \& {Pascucci}, I. 2017, Royal Society Open Science, 4, 170114

\bibitem[{{Fang} {et~al.}(2018){Fang}, {Pascucci}, {Edwards}, {Gorti},
  {Banzatti}, {Flock}, {Hartigan}, {Herczeg}, \& {Dupree}}]{Fang2018}
{Fang}, M., {Pascucci}, I., {Edwards}, S., {et~al.} 2018, \apj, 868, 28

\bibitem[{{Flock} {et~al.}(2016){Flock}, {Fromang}, {Turner}, \&
  {Benisty}}]{Mario_2016}
{Flock}, M., {Fromang}, S., {Turner}, N.~J., \& {Benisty}, M. 2016, \apj, 827,
  144

\bibitem[{{Flock} {et~al.}(2019){Flock}, {Turner}, {Mulders}, {Hasegawa},
  {Nelson}, \& {Bitsch}}]{Flock_2019}
{Flock}, M., {Turner}, N.~J., {Mulders}, G.~D., {et~al.} 2019, \aap, 630, A147

\bibitem[{{Flock} {et~al.}(2020){Flock}, {Turner}, {Nelson}, {Lyra}, {Manger},
  \& {Klahr}}]{Flock_2020}
{Flock}, M., {Turner}, N.~J., {Nelson}, R.~P., {et~al.} 2020, arXiv e-prints,
  arXiv:2005.11974

\bibitem[{{Fricke}(1968)}]{Fricke_1968}
{Fricke}, K. 1968, \zap, 68, 317

\bibitem[{{Goldreich, P. and Schubert, G.}(1967)}]{Goldreich_1967}
{Goldreich, P. and Schubert, G.} 1967, \apj, 150, 571

\bibitem[{{Gorti} {et~al.}(2009){Gorti}, {Dullemond}, \&
  {Hollenbach}}]{Gorti_2009}
{Gorti}, U., {Dullemond}, C.~P., \& {Hollenbach}, D. 2009, \apj, 705, 1237

\bibitem[{Gorti \& Hollenbach(2008)}]{Gorti_2008}
Gorti, U. \& Hollenbach, D. 2008, The Astrophysical Journal, 683, 287

\bibitem[{{Gressel} {et~al.}(2020){Gressel}, {Ramsey}, {Brinch}, {Nelson},
  {Turner}, \& {Bruderer}}]{Gressel_2020}
{Gressel}, O., {Ramsey}, J.~P., {Brinch}, C., {et~al.} 2020, arXiv e-prints,
  arXiv:2005.03431

\bibitem[{{Hollenbach} {et~al.}(1994){Hollenbach}, {Johnstone}, {Lizano}, \&
  {Shu}}]{Hollenbach_1994}
{Hollenbach}, D., {Johnstone}, D., {Lizano}, S., \& {Shu}, F. 1994, \apj, 428,
  654

\bibitem[{Hunter(2007)}]{Hunter_2007}
Hunter, J.~D. 2007, Computing in Science \& Engineering, 9, 90

\bibitem[{{Klahr} {et~al.}(2018){Klahr}, {Pfeil}, \&
  {Schreiber}}]{2018haex.bookE.138K}
{Klahr}, H., {Pfeil}, T., \& {Schreiber}, A. 2018, {Instabilities and Flow
  Structures in Protoplanetary Disks: Setting the Stage for Planetesimal
  Formation}, 138

\bibitem[{{Lin}(2019)}]{Lin2019}
{Lin}, M.-K. 2019, \mnras, 485, 5221

\bibitem[{Lin \& Youdin(2015)}]{Lin_2015}
Lin, M.-K. \& Youdin, A.~N. 2015, The Astrophysical Journal, 811, 17

\bibitem[{{Lyra} \& {Umurhan}(2019)}]{Lyra2019}
{Lyra}, W. \& {Umurhan}, O.~M. 2019, \pasp, 131, 072001

\bibitem[{{Manger} \& {Klahr}(2018)}]{Manger2018}
{Manger}, N. \& {Klahr}, H. 2018, \mnras, 480, 2125

\bibitem[{Mignone(2007)}]{Mignone_2007}
Mignone, A. 2007, \apj, 170, 228

\bibitem[{Mignone(2014)}]{Mignone_2014}
Mignone, A. 2014, Comput. Phys., 270, 784

\bibitem[{Nakatani {et~al.}(2018{\natexlab{a}})Nakatani, Hosokawa, Yoshida,
  Nomura, \& Kuiper}]{Nakatani_2018a}
Nakatani, R., Hosokawa, T., Yoshida, N., Nomura, H., \& Kuiper, R.
  2018{\natexlab{a}}, The Astrophysical Journal, 857, 57

\bibitem[{Nakatani {et~al.}(2018{\natexlab{b}})Nakatani, Hosokawa, Yoshida,
  Nomura, \& Kuiper}]{Nakatani_2018b}
Nakatani, R., Hosokawa, T., Yoshida, N., Nomura, H., \& Kuiper, R.
  2018{\natexlab{b}}, The Astrophysical Journal, 865, 75

\bibitem[{{Nelson, R.~P. and Gressel, O. and Umurhan,
  O.~M.}(2013)}]{Nelson_2013}
{Nelson, R.~P. and Gressel, O. and Umurhan, O.~M.} 2013, \aap, 435, 2610

\bibitem[{{Ormel} {et~al.}(2015){Ormel}, {Shi}, \&
  {Kuiper}}]{2015MNRAS.447.3512O}
{Ormel}, C.~W., {Shi}, J.-M., \& {Kuiper}, R. 2015, \mnras, 447, 3512

\bibitem[{{Owen} {et~al.}(2012){Owen}, {Clarke}, \& {Ercolano}}]{Owen_2012}
{Owen}, J.~E., {Clarke}, C.~J., \& {Ercolano}, B. 2012, \mnras, 422, 1880

\bibitem[{{Owen} {et~al.}(2011){Owen}, {Ercolano}, \& {Clarke}}]{Owen_2011}
{Owen}, J.~E., {Ercolano}, B., \& {Clarke}, C.~J. 2011, \mnras, 412, 13

\bibitem[{{Owen} {et~al.}(2010){Owen}, {Ercolano}, {Clarke}, \& {Alexand
  er}}]{Owen_2010}
{Owen}, J.~E., {Ercolano}, B., {Clarke}, C.~J., \& {Alexand er}, R.~D. 2010,
  \mnras, 401, 1415

\bibitem[{{Pascucci} {et~al.}(2011){Pascucci}, {Sterzik}, {Alexander},
  {Alencar}, {Gorti}, {Hollenbach}, {Owen}, {Ercolano}, \&
  {Edwards}}]{Pascucci_2011}
{Pascucci}, I., {Sterzik}, M., {Alexander}, R.~D., {et~al.} 2011, \apj, 736, 13

\bibitem[{{Pfeil} \& {Klahr}(2019)}]{Pfeil2019}
{Pfeil}, T. \& {Klahr}, H. 2019, \apj, 871, 150

\bibitem[{{Pfeil} \& {Klahr}(2020)}]{Pfeil_2020}
{Pfeil}, T. \& {Klahr}, H. 2020, arXiv e-prints, arXiv:2008.11195

\bibitem[{{Richard, Samuel and Nelson Richard P., and Umurhan Orkan
  M.}(2016)}]{Richard_2016}
{Richard, Samuel and Nelson Richard P., and Umurhan Orkan M.} 2016, \mnras,
  456, 3571

\bibitem[{{R\"udiger, G.} {et~al.}(2002){R\"udiger, G.}, {Arlt, R.}, \&
  {Shalybkov, D.}}]{rudiger_2002}
{R\"udiger, G.}, {Arlt, R.}, \& {Shalybkov, D.} 2002, A\&A, 391, 781

\bibitem[{{Sch{\"a}fer} {et~al.}(2020){Sch{\"a}fer}, {Johansen}, \&
  {Banerjee}}]{2020A&A...635A.190S}
{Sch{\"a}fer}, U., {Johansen}, A., \& {Banerjee}, R. 2020, \aap, 635, A190

\bibitem[{{Shu} {et~al.}(1994){Shu}, {Najita}, {Ostriker}, {Wilkin}, {Ruden},
  \& {Lizano}}]{Shu_1994}
{Shu}, F., {Najita}, J., {Ostriker}, E., {et~al.} 1994, \apj, 429, 781

\bibitem[{{Simon} {et~al.}(2016){Simon}, {Pascucci}, {Edwards}, {Feng},
  {Gorti}, {Hollenbach}, {Rigliaco}, \& {Keane}}]{2Simon_2016}
{Simon}, M.~N., {Pascucci}, I., {Edwards}, S., {et~al.} 2016, \apj, 831, 169

\bibitem[{{Stoll} \& {Kley}(2014)}]{2014A&A...572A..77S}
{Stoll}, M. H.~R. \& {Kley}, W. 2014, \aap, 572, A77

\bibitem[{{Tanaka} {et~al.}(2013){Tanaka}, {Nakamoto}, \&
  {Omukai}}]{Tanaka_2013}
{Tanaka}, K. E.~I., {Nakamoto}, T., \& {Omukai}, K. 2013, \apj, 773, 155

\bibitem[{Turner {et~al.}(2014)Turner, Fromang, Gammie, Klahr, Lesur, Wardle,
  \& Bai}]{Turner_2014}
Turner, N., Fromang, S., Gammie, C., {et~al.} 2014, in Protostars and planets
  VI, ed. H.~Beuther, R.~Klessen, C.~Dullemond, \& T.~Henning, Space science
  series (United States: University of Arizona Press), 411--432

\bibitem[{{Wang} {et~al.}(2019){Wang}, {Bai}, \& {Goodman}}]{Wang_2019}
{Wang}, L., {Bai}, X.-N., \& {Goodman}, J. 2019, \apj, 874, 90

\end{thebibliography}

\end{document}